\documentclass[aps,amsmath,amssymb,prl,superscriptaddress,twocolumn]{revtex4}

\usepackage{epsfig}
\usepackage{amsmath}
\usepackage{amssymb}
\begin{document}

\title{Dependence on temperature and GC content of bubble length
distributions in DNA}


\author{G. Kalosakas$^*$}
\affiliation{Department of Materials Science, University of Patras,
GR-26504 Rio, Greece}
\author{S. Ares\footnote{Both authors contributed equally to this work.}}
\affiliation{Max Planck Institute for the Physics of Complex
Systems, N\"othnitzer Str. 38, D-01187 Dresden, Germany,
and Grupo Interdisciplinar de Sistemas Complejos (GISC)}




\begin{abstract}
We report numerical results on the temperature dependence
of the distribution of bubble lengths in DNA segments
of various GC concentrations. Base-pair openings are described
by the Peyrard-Bishop-Dauxois model and the corresponding thermal
equilibrium distributions of bubbles are obtained through
Monte Carlo calculations for bubble sizes up to the order of a hundred
base-pairs. The dependence of the parameters of bubble length
distribution on temperature and the GC content is investigated.
We provide simple expressions which approximately describe
these relations. The variation of the average bubble length is also
presented.
Finally, we find a temperature dependence of the exponent $c$ in the
probability distribution of bubble lengths. If an analogous dependence
exists in the loop entropy exponent of real DNA, it may be
relevant to understand overstretching in force-extension experiments.
\end{abstract}

\maketitle

\section{I. Introduction}

Local openings of the DNA double helix are required in several biological
functions, for instance transcription and replication. These local
separations of the two DNA strands are mediated by a specific
machinery in the cell. In order to deal with such complex processes, it
is necessary first to understand the interactions keeping together
the two complementary strands within a single DNA duplex, as well as
the properties of fluctuating DNA openings in thermal equilibrium.
This fact has been realized long time ago and base-pairing interactions
and the thermal stability of the double helix
have been conventionally probed by increasing temperature up to the DNA
denaturation transition \cite{WB,PS}.

Base-pair openings (bubbles) occur in DNA due to thermal
fluctuations even at temperatures well below the melting transition.
It has been
speculated that they may play a role in the recognition of specific
DNA sites by DNA-binding proteins \cite{sobell,NARa,NARb,BiophJ}.
Recent experiments
using a novel hairpin quenching technique have been able to show
this bubble formation in the pre-melting regime and characterize
it quantitatively \cite{zocchi,zocchi3,zocchi2}.
By increasing
temperature these bubbles grow and more bubbles are nucleated,
thus leading to the complete separation of the two strands at the
denaturation transition.
Therefore statistical properties of DNA bubbles in a wide temperature
regime, extending from biological temperatures up to the melting
transition, are of particular interest. The purpose of this work is the
investigation of the distribution of bubble lengths and its temperature
dependence for DNA sequences containing different percentages of
guanine-cytosine (GC) base-pairs. The study is performed in the
framework of the Peyrard-Bishop-Dauxois (PBD) model \cite{DPB},
from where conjectures are drawn about the actual distributions
of bubbles in DNA.

If our results on the temperature dependence of this distribution,
obtained for finite bubble lengths, remain qualitatively valid in the
asymptotic regime (for very large values of bubble lengths), then there 
would be a connection with the ongoing discussion about the interpretation
of the DNA overstretching observed in force-extension experiments.
In this case the possibilities of either a force-induced melting
or the existence of a double-stranded elongated DNA phase (the
so called S-DNA) have been proposed to explain the abrupt
elongation of DNA at forces $\sim65 pN$ \cite{williams}.
This is briefly discussed in the concluding section.

In a recent study we have presented the distribution of bubble lengths
in the PBD model of DNA at $310$K and we found that it can be described
by a power-law modified exponential \cite{AK}.
Anharmonic interactions between
complementary bases forming base-pairs are responsible for the observed
non-exponential distribution. The same form of distribution
has been derived in the framework of the Poland-Scheraga model
\cite{poland,kafri,coluzzi}, which represents a completely
different theoretical approach of DNA denaturation than the
PBD model that we use in our calculations.
This distribution is also found for a primitive version
of the PBD model, viz. the Peyrard-Bishop model \cite{PB} with
linear stacking interactions, but in this case the characteristic values
of the parameters of the distribution are different \cite{sung}.

Here we examine how the bubble length distribution varies with
temperature and present its complete dependence on both temperature
and the GC fraction of the DNA segment. The PBD model \cite{DPB}
is used for the description of base-pair openings, where a set of
continuous variables $y_n$ represent the base-pair displacements from
equilibrium distance and the index $n$ labels the base-pairs along
the DNA chain. The potential energy of the system consists of two parts:
the on-site interaction $V(y_n)$ within each base-pair and the
stacking interaction $U(y_n,y_{n+1})$ between adjacent base-pairs.
A Morse potential is used for the on-site energy,
\begin{equation}   \label{on-site}
V(y_n)=D_n(e^{-a_ny_n}-1)^2,
\end{equation}
where the parameters $D_n$ and $a_n$
distinguish between GC and AT base-pairs ($D_{GC}=0.075 $eV,
$a_{GC}=6.9 $\AA$^{-1}$ for a GC base-pair and
$D_{AT}=0.05 $eV, $a_{AT}=4.2 $\AA$^{-1}$ for an AT pair), while
a nonlinear potential describes the stacking interaction,
\begin{equation}   \label{stacking}
U(y_n,y_{n+1})=\frac{K}{2}(1+\rho e^{-b (y_n+y_{n+1})})(y_n-y_{n+1})^2,
\end{equation}
with $K=0.025 $eV/\AA$^2$, $\rho=2$, and $b=0.35 $\AA$^{-1}$.
We use parameter values from previous works
\cite{NARa,NARb,CG,saul}. These parameters have been originally obtained
from empirical fits to experimental data \cite{CG} and are also able
to successfully describe other experimental situations \cite{NARa,saul}.
Entropic effects of DNA are described
from the PBD Hamiltonian when studying its thermodynamics, leading
to an entropy driven melting transition \cite{DPB,DP}.

The efficiency of the rather simple PBD model to describe base-pair
openings in DNA has led to its extensive use in the literature
\cite{DP,CH,CG,TDP,NARa,NARb,VKRB,saul,vanerp,rapti,krastan,cpl,VRBR,AK,theod,SHM,das,BiophJ,alexandrov09}.
In particular, our choice of the PBD model for the study of bubble length
distributions is motivated by the success of the model \cite{saul}
in reproducing experimental measurements of bubble
formation \cite{zocchi,zocchi3}, as well as the accurate description
at a quantitative level of melting curves in short DNA segments \cite{CG}.
The coarse-grained description of the model allows the possibility to
perform calculations with long sequences of up to tens of thousands
of base pairs \cite{SHM}.
Moreover, the popular Nearest Neighbor model \cite{santalucia},
which is very successful in describing the melting of short oligomers,
does not reproduce well experimental data on intermediate states for
longer sequences \cite{gonzalez,EKS,zocchi3,zocchi2}.
Compared to the more conventional modified
Ising type models used in the study of DNA melting \cite{poland,PS},
the PBD model is qualitatively similar: the on-site potential of the PBD
model is an extension of the {\em magnetic field} in the Ising type models,
while the nonlinear stacking potential in Eq. (\ref{stacking}) corresponds
to the interaction between neighboring spins and loop entropies in Ising
models. We expect our results for the PBD model to hold qualitatively
also for the Ising type models, since both models yield the same shape
for the distribution of bubble sizes \cite{AK}.

\section{II. Results}

Considering a random DNA sequence of a given GC percentage, $x_{GC}$, at
equilibrium at temperature $T$, we calculate the distribution per base-pair
of bubble lengths $l$, $P(l)$,
by counting during Monte Carlo simulations of the DPB model the average
occurrences of openings (base-pair displacements) larger than a fixed
threshold $y_{thres}$ at $l$ successive base-pairs. In each simulation
a random sequence of 1000 base pairs is used in which the AT or GC
identity of each base pair is generated randomly under the
constraint of a specified GC percentage. The properties of such random
long sequences are not different from natural sequences, as we showed
in previous work \cite{AK} comparing the results for a segment from
{\em Escherichia Coli}'s {\em gal} promoter with those from a
randomly generated sequence with the same GC percentage.

The Monte Carlo simulation is performed using the
Metropolis algorithm \cite{metropolis}, which is used first
for thermalization and then for measurement runs. Results
are averaged over several realizations using the same random sequence and
different initial conditions for the inter-base distance within each
base-pair and for the random number generator. Other details of the
simulation (number of realizations and Monte Carlo steps) are as in
Ref. \cite{AK}. Since the used sequences are so
long, the results are independent of the precise realization of the random
sequence. We have run further simulations with different random sequences to
assure this point. There is an issue about the dissociation observed
in the PBD model in Monte Carlo simulations. In particular, due to the
upper bound of the on-site potential for large positive displacements,
for long enough simulations a complete dissociation would eventually
occur at any temperature, even below the melting transition
\cite{DP,zhang}. However, because of the large DNA segments considered
here, the probability of a complete dissociation during the
simulation time is so low that no such events were observed
at any of the studied temperatures below the melting temperature.

The threshold value considered for the openings is $y_{thres}=1.5 \AA$.
Results for different values of this threshold are qualitatively similar.
We use periodic boundary conditions
in DNA segments of length 1000 base-pairs (sufficiently larger than
the studied bubble sizes) and therefore our results refer to internal
bubbles in long DNA chains. The results are independent of the
length of the studied sequence, provided that it is longer than the
longest observed bubbles \cite{AK}. This has been verified by simulations
on DNA segments of different length,
resulting in the same distributions $P(l)$ for the bubble sizes studied
here, i.e. with $l$ up to about a hundred base-pairs.
For short molecules, the actual sequence
does play a role, since it can trigger local openings in different
regions at the end or in the middle of the sequence \cite{zocchi,zocchi3}.
In this case, the border effects associated can not be characterized only by
the GC percentage of the short sequence, but this problem is out of the
scope of the present work.

\begin{figure}
\includegraphics[width=8.0cm]{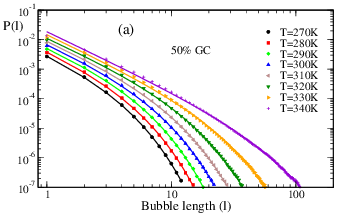} \\ 
\includegraphics[width=8.0cm]{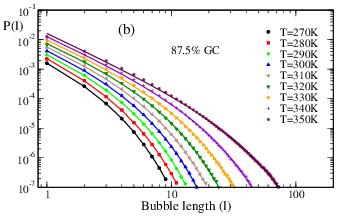} \\ 
\caption{\label{fig:1} (Color online) Distribution per base-pair of bubble lengths $l$ (in
number of base-pairs), $P(l)$, for different temperatures (points, as
indicated in the plots) at random DNA sequences with a GC content of
(a) 50 \% and (b) 87.5 \%. Continuous lines are fits with the
distribution of Eq.(\ref{dis}). $y_{thres}=1.5$\AA.
}
\end{figure}

In Figure \ref{fig:1} we show bubble length distributions at various
temperatures for a GC percentage of $x_{GC}=$ 50\% and 87.5\%.
Similar plots have been obtained for nine different values
of $x_{GC}$: 0\%,12.5\%,25\%,37.5\%,$\ldots$,100\%.
Note here that although the PBD model takes into account cooperativity in
bubble formation through the nonlinear stacking interaction, it does not
contain any parameter describing a bubble nucleation size. Thus the
distributions in figure \ref{fig:1} show results even for bubbles of length
$l=1$, implying that the nucleation size is 1. This should not be confused
with the minimum length of a short DNA sequence necessary to sustain bubble
states, which has been shown experimentally \cite{zocchi,zocchi3} and
theoretically within the DPB model \cite{saul} to be greater than one.

Except for the cases of
$l=1$ at relatively higher temperatures (closer to the melting transition),
the numerical results obtained for the distributions can be rather well
described by the power-law modified exponential \cite{AK}
\begin{equation}   \label{dis}
P(l) = W \frac{e^{-l/\xi}}{l \; ^c}, \hspace{0.5cm} \mbox{for} \; l> 1.
\end{equation}
In the following we characterize the dependence of the parameters of
distribution on $T$ and $x_{GC}$, and provide approximate expressions
for the relations $\xi(T,x_{GC})$, $c(T,x_{GC})$, and $W(T,x_{GC})$.
The distribution parameters are obtained through fitting of plots
like those of Figure \ref{fig:1} with Eq. (\ref{dis}), using a weight
proportional to $1/P(l)^2$.

We note here that Eq. (\ref{dis}) can be derived from polymer physics
\cite{poland,kafri,coluzzi} as an {\it asymptotic} expression for very
large bubble sizes, $l \gg 1$, in DNA. However, we find that the same
expression describes well the bubble length distribution in the PBD model
of DNA, even for small bubble sizes $l$. We emphasize that we are interested
here in bubble lengths up to about a hundred, or a few hundred at most,
base-pairs (which are relevant for any practical purpose) and {\it not} for
the asymptotic behavior of the distribution. In this context we use
Eq. (\ref{dis}) as an empirical formula valid for finite bubble lengths
$l$ (in the non-asymptotic regime), and describe the variation of its
parameters on $T$ and $x_{GC}$. If one is interested in the asymptotic
behavior of the distribution, this can be obtained by proper scaling
analysis and, as it has been shown in Ref. \cite{theod}, it may be described
by different values of parameter $c$. Therefore we are not
concerned with the order of the melting transition and the exponent $c$
presented here is not indicative of the kind of the transition
\cite{TDP,kafri,coluzzi,theod,EKS}, as it also depends on the somehow
arbitrary value of the threshold chosen to consider a base-pair open.

\begin{figure}
\includegraphics[width=8.0cm]{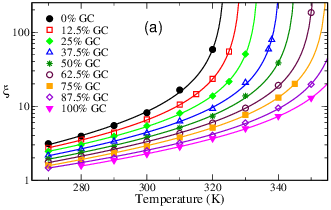} \\ [1mm]
\includegraphics[width=8.0cm]{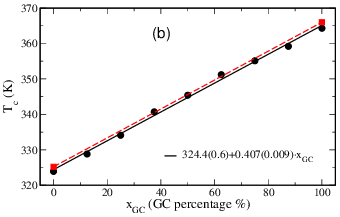} \\ [3mm]
\includegraphics[width=8.0cm]{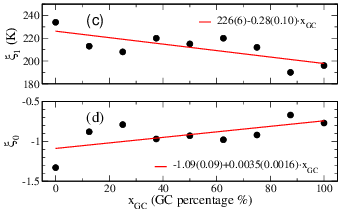} \\ 
\caption{\label{fig:2} (Color online) (a) Dependence of the decay length $\xi$
of the distribution (\ref{dis}) on the temperature, for different values
of the GC content of the DNA sequence (points).
Lines show fits with the function of Eq. (\ref{xiT}).
(b) Dependence of the critical temperature $T_c$, as obtained from
the fitting of the $\xi(T)$ data with Eq. (\ref{xiT}), on the GC content
of the DNA sequence (circles). Solid line represents a least square fit
according to a linear dependence. Squares show exact results of the critical
temperatures for the homogeneous cases of $0 \%$ GC and $100 \%$ GC, obtained
from transfer integral calculations, while the dashed line connects these
two points.
(c) and (d) Dependence of the parameters $\xi_1$ and $\xi_0$,
respectively, of the fit of the $\xi(T)$ data with Eq. (\ref{xiT}),
on the GC content of the sequence (circles). Solid lines represent
linear fits of the corresponding data. Equations of straight lines
resulting from the corresponding fittings are shown in (b), (c), and (d),
where the values in parentheses represent errors of the fitting parameters.}
\end{figure}

Figure \ref{fig:2} presents the dependence of the decay length $\xi$
of Eq. (\ref{dis}). In \ref{fig:2}a the variation of $\xi$ with
temperature is shown (points) for different values of $x_{GC}$. The
T-dependence is accurately described by the divergent function
\begin{equation}   \label{xiT}
\xi(T) = \xi_0 + \frac{\xi_1}{T_c-T},
\end{equation}
where $T_c$ is the denaturation temperature. Such a relation is also valid
for the Poland-Scheraga model \cite{coluzzi}. Lines in Figure \ref{fig:2}a
show fittings of the $\xi(T)$ data with Eq. (\ref{xiT}), using a
weight proportional to $1/\xi^2$. The critical
temperature $T_c$ as obtained from the fitting at different values
of $x_{GC}$ is presented with circles in Figure \ref{fig:2}b.
A linear dependence of $T_c$ on the GC content is found, in accordance
with known experimental results \cite{MD} and calculations from simplified
models \cite{asepjb}. A least square fitting of the
$T_c(x_{GC})$ data results in the continuous line shown in Figure \ref{fig:2}b.
Regarding the homopolymer cases of poly(dA)-poly(dT) ($x_{GC}=0\%$) and
poly(dG)-poly(dC) ($x_{GC}=100\%$), the critical temperatures for the
transition can be independently calculated through the numerically exact
transfer integral technique \cite{ti1,ti2,bsg}, and the corresponding results
are $T_c(x_{GC}=0)=325.2$K and $T_c(x_{GC}=100)=366.0$K. These values
are shown with squares joined by a dashed line in Figure \ref{fig:2}b,
where the line lies inside the error interval of the Monte Carlo results.
The actual denaturation temperatures obtained through the Monte Carlo
simulations are in agreement with those presented in Fig. \ref{fig:2}b.
Therefore, the critical temperatures $T_c$ shown in Fig. \ref{fig:2}b
provide a guide of how far from the melting transition are the data
at various temperatures presented in this work.

The other parameters $\xi_1$ and $\xi_0$ resulting from the fitting of the
$\xi(T)$ data with Eq. (\ref{xiT}) are shown with circles in Figures
\ref{fig:2}c and \ref{fig:2}d, respectively. Their dependence on
$x_{GC}$ can be approximately considered as linear (continuous lines
in Figs. \ref{fig:2}c and \ref{fig:2}d). Therefore, the relation
\begin{equation}   \label{xiTgc}
\xi(T,x_{GC}) = a_1 + a_2 x_{GC} + \frac{a_3+a_4 x_{GC}}{a_5+a_6 x_{GC}-T} \; ,
\end{equation}
where $a_1,a_2,\ldots,a_6$ are constants, can approximately provide
the dependence of the decay length $\xi$ on $T$ and $x_{GC}$.

\begin{figure}
\includegraphics[width=8.0cm]{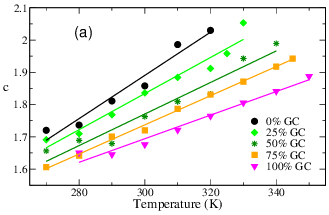} \\ [1mm]
\includegraphics[width=8.0cm]{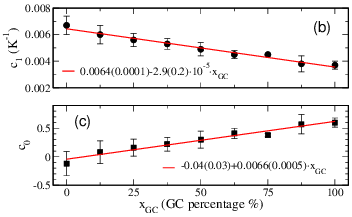} \\ 
\caption{\label{fig:3} (Color online) (a) Dependence of the exponent $c$
of the distribution (\ref{dis}) on the temperature, for different values
of the GC content of the DNA sequence (points).
Lines show linear fits with equation (\ref{cT}).
(b) and (c) Dependence of the parameters $c_1$ and $c_0$,
respectively, of the fit of the $c(T)$ data with Eq. (\ref{cT}),
on the GC content of the sequence (filled points). Error bars are standard
errors resulting from the fitting procedure. Solid lines represent linear
fits of the corresponding data and the resulting equations are also shown.}
\end{figure}

The variation of the exponent $c$ of the distribution (\ref{dis}) is
presented in Figure \ref{fig:3}. Points in \ref{fig:3}a show the
temperature dependence of $c$ for different GC percentages (for clarity
of the plot the corresponding results for
$x_{GC}=12.5\%,37.5\%,62.5\%,87.5\%$ have been omitted).
This dependence may approximately  be described by a linear function
\begin{equation}   \label{cT} 
c(T) = c_0 + c_1 T. 
\end{equation}
Lines in Figure \ref{fig:3}a show fittings of the numerical results
with the above formula. The parameters $c_1$ and $c_0$ obtained from the
fitting at different values of GC content are shown with points in
Figures \ref{fig:3}b and \ref{fig:3}c, respectively, while the corresponding
error bars are derived from the fitting procedure. The latter plots indicate
a linear dependence of $c_1$ and $c_0$ on $x_{GC}$, implying the
approximate relation
\begin{equation}   \label{cTgc}
c(T,x_{GC}) = b_1 + b_2  x_{GC} + (b_3+b_4  x_{GC})  T ,
\end{equation}
where $b_1$, $b_2$, $b_3$, and $b_4$ are constants independent of
$T$ and $x_{GC}$.

\begin{figure}
\includegraphics[width=8.0cm]{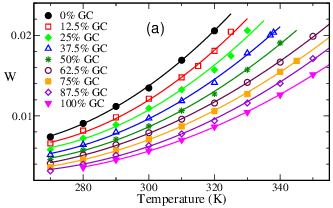} \\ [2mm]
\includegraphics[width=8.0cm]{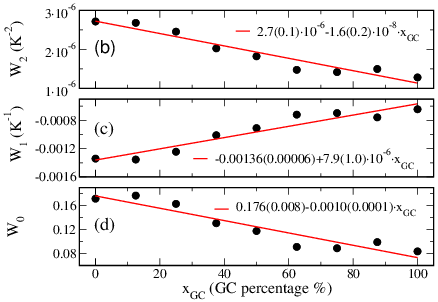} \\ 
\caption{\label{fig:4} (Color online) (a) Dependence of the pre-exponential coefficient
$W$ of the distribution (\ref{dis}) on the temperature, for different values
of the GC content of the DNA sequence (points).
Lines show quadratic fits with equation (\ref{WT}).
(b), (c), and (d) Dependence of the parameters $W_2$,
$W_1$, and $W_0$, respectively, of the fit of the $W(T)$ data with
Eq. (\ref{WT}), on the GC content of the sequence (circles).
Solid lines represent linear fits of the corresponding data and the
resulting equations are also shown.}
\end{figure}

In Figure \ref{fig:4} is shown the dependence of the coefficient $W$
of the distribution (\ref{dis}). Points in \ref{fig:4}a display the
variation of $W$ with $T$ for various GC contents. A quadratic function
\begin{equation}   \label{WT}
W(T) = W_0 + W_1 T + W_2  T^2
\end{equation}
can describe rather well the temperature dependence of $W$, at least
in the studied temperature regime. The corresponding fittings with
Eq. (\ref{WT}) are shown by lines in Figure \ref{fig:4}a. The resulting
fitting parameters at different GC percentages are plotted by points
in Figures \ref{fig:4}b, \ref{fig:4}c, and \ref{fig:4}d, respectively.
These plots can approximately be described by linear dependences of
$W_2$, $W_1$, and $W_0$ on $x_{GC}$ (solid lines).
As a result the expression
\begin{equation}   \label{WTgc}
W(T,x_{GC}) = d_1 + d_2 x_{GC} + (d_3+d_4 x_{GC}) T
+ (d_5 + d_6  x_{GC}) T^2
\end{equation}
can approximate the dependence of $W$ on $T$ and $x_{GC}$, where
$d_1,d_2,\ldots,d_6$ are constants.

The detailed results of this investigation do not confirm a bilinear
dependence of the parameters of the distribution on the GC content
at a fixed temperature \cite{AK}.
Instead, a linear dependence of the exponent
$c$ and the coefficient $W$ on $x_{GC}$ arises at constant $T$.
We note that the results presented in Figure 2 of Ref. \cite{AK} can
be described well by Eqs. (\ref{xiTgc}), (\ref{cTgc}), and (\ref{WTgc})
for $T=310$K and using the values of constants as they provided
in Figures \ref{fig:2}b,c,d (for $a_i$), \ref{fig:3}b,c (for $b_i$),
and \ref{fig:4}b,c,d (for $d_i$).

We finally present the variation of the average
bubble length, $L_B$, on the two parameters of interest,
viz. temperature and GC percentage. $L_B$ is
obtained through the total number of base-pairs in bubble states
divided by the total number of bubbles \cite{AK}:
\begin{equation} \label{avl}
L_B=\frac{\sum_l lP(l)}{\sum_{l \geq 1} P(l)}.
\end{equation}
Substituting $P(l)$ by the power-law modified exponential function
of Eq.~(\ref{dis}), the sums in Eq.~(\ref{avl}) yield:
\begin{equation} \label{avls}
L_B(c,\xi)=\frac{\text{Li}_{c-1}(e^{-1/\xi})}{\text{Li}_{c}(e^{-1/\xi})},
\end{equation}
where Li$_s(z)$ is the first branch of the polylogarithm function
\cite{lewin}, defined as Li$_s(z)=\sum_{k=1}^\infty z^k/k^s$.
The average bubble length for a given temperature and GC percentage,
$L_B(T,x_{GC})$, can be obtained substituting in Eq.~(\ref{avls}) the
parameters $\xi$ and $c$, calculated for the corresponding $T$ and $x_{GC}$
through Eqs.~(\ref{xiTgc}) and (\ref{cTgc}), respectively, using the values
of constants determined in Figures~\ref{fig:2} and~\ref{fig:3}.

\begin{figure}
\includegraphics[width=8.0cm]{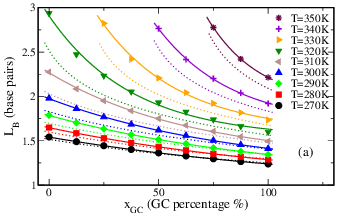} \\ 
\includegraphics[width=8.0cm]{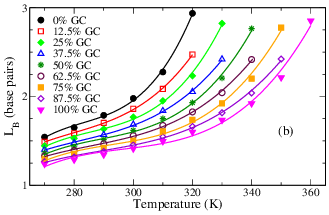} \\ 
\caption{\label{fig:5} (Color online) (a) Dependence of the average bubble length
$L_B$, Eq. (\ref{avl}), on the GC content, for different temperatures (points).
Dotted lines show the analytical result of Eq.~(\ref{avls}), using the values
of $\xi$ and $c$ from Eqs. (\ref{xiTgc}) and (\ref{cTgc}).  Solid lines show
empirical fits with exponential functions, Eq.~(\ref{avled}). (b)
Dependence of the average bubble length $L_B$ on temperature for different
GC contents (points). Lines represent empirical fits with cubic functions.}
\end{figure}

Figure \ref{fig:5}a depicts the dependence of $L_B$ on $x_{GC}$ for
different temperatures, obtained directly from our numerical simulations
(points). Eq.~(\ref{avls}), shown by dotted lines in the plot,
reproduces correctly the behavior of the data.
However, the errors in the estimated parameters
in Figures~\ref{fig:2} and~\ref{fig:3} add up to produce a small
deviation between the numerical points and the values obtained by
Eq.~(\ref{avls}). We also show, using solid lines in the plot, more
accurate fits of the numerical data with a simpler phenomenological
expression, namely the exponentially decaying function
\begin{equation} \label{avled}
L_B = \Gamma_0 + \Gamma_1 \exp(-\Gamma_2 x_{GC}).
\end{equation}
These empirical fits generalize the apparent exponential decay seen earlier
at $310K$ \cite{AK}. Presenting these data on a different way, more
appropriate for experimental investigations, the temperature dependence of
$L_B$ for fixed GC contents is shown in figure \ref{fig:5}b (points), for
temperatures sufficiently below the melting transition. Here for clarity of
the plot we do not show again the prediction of Eq.~(\ref{avls}), but instead
we show with lines fits with a simple phenomenological function, namely a
cubic polynomial, which describes approximately the data in the investigated
temperature regime. The coefficients of the cubic polynomials show an
exponential dependence, of the form (\ref{avled}), on the GC percentage.

\section{III. Conclusions}

In conclusion, we have presented the dependence of bubble length
distributions in the PBD model of DNA on temperature and the GC content.
The investigated temperature regime was extended from biologically
relevant values up to values below the melting transition.
Approximate expressions have been obtained for the parameters
of the power-law modified exponential distribution in the non-asymptotic
regime (for bubble lengths up to about a hundred base-pairs).
The exponent $c$ behaves linearly both in temperature and GC content,
Eq. (\ref{cTgc}), while the coefficient $W$ shows a quadratic
dependence on temperature and a linear dependence on the GC fraction,
Eq. (\ref{WTgc}). The decay length $\xi$ is described by the
relatively simple equation (\ref{xiTgc}). The constants $a_i$, $b_i$,
and $d_i$ appearing in these expressions depend on the amplitude
$y_{thres}$ of the considered base-pair openings, and for
$y_{thres}=1.5${\AA} are given by the values shown in Figures
\ref{fig:2}, \ref{fig:3}, and \ref{fig:4}. Using the expressions of the
exponent $c$ and the decay length $\xi$, the average bubble length
is analytically given by Eq.~(\ref{avls}).
Our results may be useful in biotechnological applications which involve
thermally induced DNA denaturation, as well as in recent
hairpin quenching experiments \cite{zocchi,zocchi3,zocchi2}
for studying bubble formation.

An important result of this work is the prediction of
a dependence both in temperature and genetic sequence of the
exponent $c$ of the bubble distribution (\ref{dis}). This prediction
can be experimentally verified using the method proposed in
Ref. \cite{bar}: based on the measurement of certain
correlation functions of base-pair openings using fluorescence
correlation spectroscopy \cite{altan}, the exponent $c$ can be
experimentally calculated.

\begin{figure}
\includegraphics[width=8.0cm]{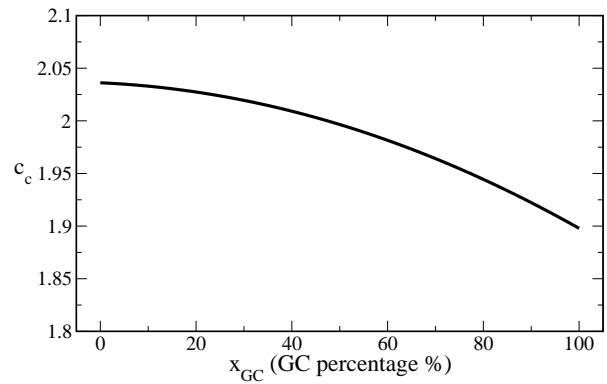} \\ 
\caption{\label{fig:6} Extrapolated dependence of the exponent $c$ at the
critical temperature ($c_c$) on the GC content.
}
\end{figure}

The sequence dependence of $c$ is specially relevant at the melting
temperature. At this temperature very large bubbles appear
and the asymptotic behavior of the bubble length distribution is necessary.
Assuming that the behavior revealed in our findings qualitatively holds
in the asymptotic regime (which needs to be investigated since there
is not {\it a priori} any reason for this to be true), then the particular
dependence of $c$ can be extrapolated. Combining an expression for
$c$ as that given in Eq.~(\ref{cTgc}) and the dependence of
critical temperature on the GC percentage presented in Fig.~\ref{fig:2}b,
we show in Fig.~\ref{fig:6} estimated values of $c$ at the
melting temperature, $c_c$, as a function of the GC percentage.
We see that $c_c$ is not constant, but decreases as the GC fraction
increases. Although, in view of the above assumptions, this result
should be taken with caution, such a behavior is consistent with the
observation, coming from lattice models simulations that consider
DNA strands as self-avoiding random walks, of
a decrease of $c_c$ with increasing stiffness of the strands \cite{carlon}.

Under the same assumption that our results for the temperature dependence
of $c$ qualitatively hold in the asymptotic regime, then this may be
useful to understand experimental results in force-extension
experiments. When performing single molecule experiments, force-extension
curves of double stranded DNA display hysteresis \cite{rief,mao}.
The hysteresis observed is more pronounced at higher temperatures.
The existence of an elongated double-stranded form of DNA, called S-DNA,
has been proposed to explain the abrupt elongation of DNA at a force
of $\sim65 pN$ \cite{sdna}.
Numerical studies of the force-extension problem  \cite{WPG}
suggest that the existence of S-DNA is necessary to explain the
temperature dependence of the observed hysteresis. However,
an alternative explanation without the appearance of an S-DNA phase
is possible
if the exponent $c$ is temperature dependent \cite{WPG}. The order of
variation of $c$ we find here is in the right direction ($c$ increases with
temperature, implying larger hysteresis loops at high temperatures according
to the experiments) and of the necessary magnitude \cite{WPG} to explain
the hysteresis effects observed. Hence, if the qualitative behavior of our
results is valid in the asymptotic regime, this would have an effect on the
ongoing debate about S-DNA, favoring force-induced
melting as an explanation for the DNA elongation over the formation
of an S-DNA phase. Interestingly, recent experimental results
support the force-induced melting explanation \cite{shokri},
in agreement with the expectation from the temperature dependence of $c$.

{\it Acknowledgments.}
We thank J. Bois for critical reading of the manuscript
and J. Bois and N. Theodorakopoulos for enlightening discussions.
G.K. acknowledges the hospitality of MPI-PKS in Dresden and
support from the C. Caratheodori program C155 of University of Patras.
S.A. acknowledges financial support from
Ministerio de Educaci\'on y Ciencia (Spain) through grant MOSAICO.

\end{document}